\documentclass[lettersize,journal]{IEEEtran}
\usepackage{amsmath,amsfonts}
\usepackage{algorithmic}
\usepackage{algorithm}
\usepackage{array}
\usepackage[caption=false,font=normalsize,labelfont=sf,textfont=sf]{subfig}
\usepackage{textcomp}
\usepackage{stfloats}
\usepackage{url}
\usepackage{verbatim}
\usepackage{graphicx}
\usepackage{cite}
\usepackage{enumitem}
\usepackage{makecell}
\begin{document}

\title{Next-Generation Wi-Fi Networks with Generative AI: Design and Insights}

\author{Jingyu Wang, Xuming Fang,~\IEEEmembership{Senior Member,~IEEE}, Dusit Niyato,~\IEEEmembership{Fellow,~IEEE}, Tie Liu
        % <-this % stops a space
\thanks{J. Wang, X. Fang and T. Liu are with Key Lab of Info Coding \& Transmission, Southwest Jiaotong University, Chengdu 610031, China. 
(E-mails: wangmr930@my.swjtu.edu.cn, xmfang@swjtu.edu.cn, tliu@my.swjtu.edu.cn).}
\thanks{D. Niyato is with the College of Computing and Data
 Science, Nanyang Technological University, Singapore 639798 (e-mail: dniyato@ntu.edu.sg).}}% <-this % stops a space

% {Shell \MakeLowercase{\textit{et al.}}: Generative AI for Wi-Fi Network}


\maketitle

\begin{abstract}
Generative artificial intelligence (GAI), known for its powerful capabilities in image and text processing, also holds significant promise for the design and performance enhancement of future wireless networks. In this article, we explore the transformative potential of GAI in next-generation Wi-Fi networks, exploiting its advanced capabilities to address key challenges and improve overall network performance. We begin by reviewing the development of major Wi-Fi generations and illustrating the challenges that future Wi-Fi networks may encounter. We then introduce typical GAI models and detail their potential capabilities in Wi-Fi network optimization, performance enhancement, and other applications. Furthermore, we present a case study wherein we propose a retrieval-augmented LLM (RA-LLM)-enabled Wi-Fi design framework that aids in problem formulation, which is subsequently solved using a generative diffusion model (GDM)-based deep reinforcement learning (DRL) framework to optimize various network parameters. Numerical results demonstrate the effectiveness of our proposed algorithm in high-density deployment scenarios. Finally, we provide some potential future research directions for GAI-assisted Wi-Fi networks.
\end{abstract}


\section{Introduction}
\IEEEPARstart{R}{ecently}, OpenAI's Chat Generative Pre-trained Transformer (ChatGPT) has garnered significant attention from academic and industrial communities, promoting the application of generative AI (GAI) models across various fields. Unlike traditional AI, which specializes in classification and prediction tasks, GAI can learn complex data distributions and generate new similar samples. Due to its powerful generating ability, GAI can handle a wide variety of tasks, including image, audio, and text processing. For instance, diffusion models have been extensively applied in image generation and image super-resolution \cite{ref1}. In addition, GAI possesses distinguished learning and inference capabilities, which is a promising technology for helping wireless network optimization to meet the demands of the iterative development of communication systems.
 
Wi-Fi networks have become a primary means of Internet access, significantly easing the traffic burden on costly cellular networks. However, the rapid popularization of extended reality (XR), along with 4k and 8k videos, poses enormous challenges to Wi-Fi's carrying capacity. To handle this issue, next-generation Wi-Fi introduces numerous novel features to fulfill ultra-high throughput and stringent low-latency requirements \cite{dengcailian}. While these new features definitely enhance Wi-Fi performance, they also introduce additional complexity, making Wi-Fi network configuration more intricate. For example, the newly introduced multi-link operation (MLO) complicates channel allocation across different links, while multi-access point (AP) coordination entails more intricate transmission setups. These innovations involve numerous adjustable parameters, necessitating careful alignment with appropriate scenarios—a process that demands extensive time for retrieval and analysis. Traditional AI methods, typically used for fixed-pattern tasks such as classification and prediction, lack dynamic adaptation capabilities and heavily rely on extensively annotated datasets, making them inadequate for modeling complex wireless systems. In contrast, GAI, thanks to its strong understanding and modeling capabilities, has immense potential to address these intractable problems.

Although there have been efforts to incorporate GAI into wireless communication, its application in Wi-Fi systems, particularly for enhancing network performance, remains limited. Most existing works focus on GAI-assisted Wi-Fi-based applications, e.g., indoor location \cite{IndoorLocalization}, with insufficient attention to improving communication performance. However, GAI can be also effectively utilized to optimize Wi-Fi network communication, addressing its complex parameters adjusting and highly dynamic wireless environment adaptation, especially in scenarios with substantial stations. For instance, GAI can analyze user data from the current wireless environment to assess channel conditions. If the channel is congested, GAI can adjust parameters such as modulation order, contention window size, and frame length to maintain reliable data transmission. Through employing GAI method in Wi-Fi network, researchers can better understand and manage the complexities of various Wi-Fi scenarios, ensuring systems are adaptable to the evolving needs of wireless communication. \textit{To the best of the authors' knowledge, this is the first work to systematically presents the application of GAI in facilitating Wi-Fi network optimization and enhancing performance, particularly in communication.} The main contributions are as follows.

\begin{itemize}
    \item{We first review the evolution of main Wi-Fi generations and discuss the primary application services of Wi-Fi networks. Following this, we illustrate the challenges that the next generation and beyond of Wi-Fi networks may face.}
    \item{We introduce the several typical GAI models and the large language model (LLM). Subsequently, we discuss applications of the LLM in assisting researchers to analyze scenario requirements and formulate problems. Finally, we explore the GAI-enhanced technologies in Wi-Fi network for performance enhancement.}
    \item{We propose a novel framework using the LLM and other assistive technologies to facilitate Wi-Fi networks design. Moreover, we present a case study on problem design for high-density deployment scenarios, and propose a generative diffusion model (GDM) method to solve the formulated problem. The results demonstrate the advantages of GAI in Wi-Fi network optimization and performance enhancement.}
\end{itemize}

\begin{figure*}[ht]
    \centering
    \includegraphics[width=0.85\textwidth]{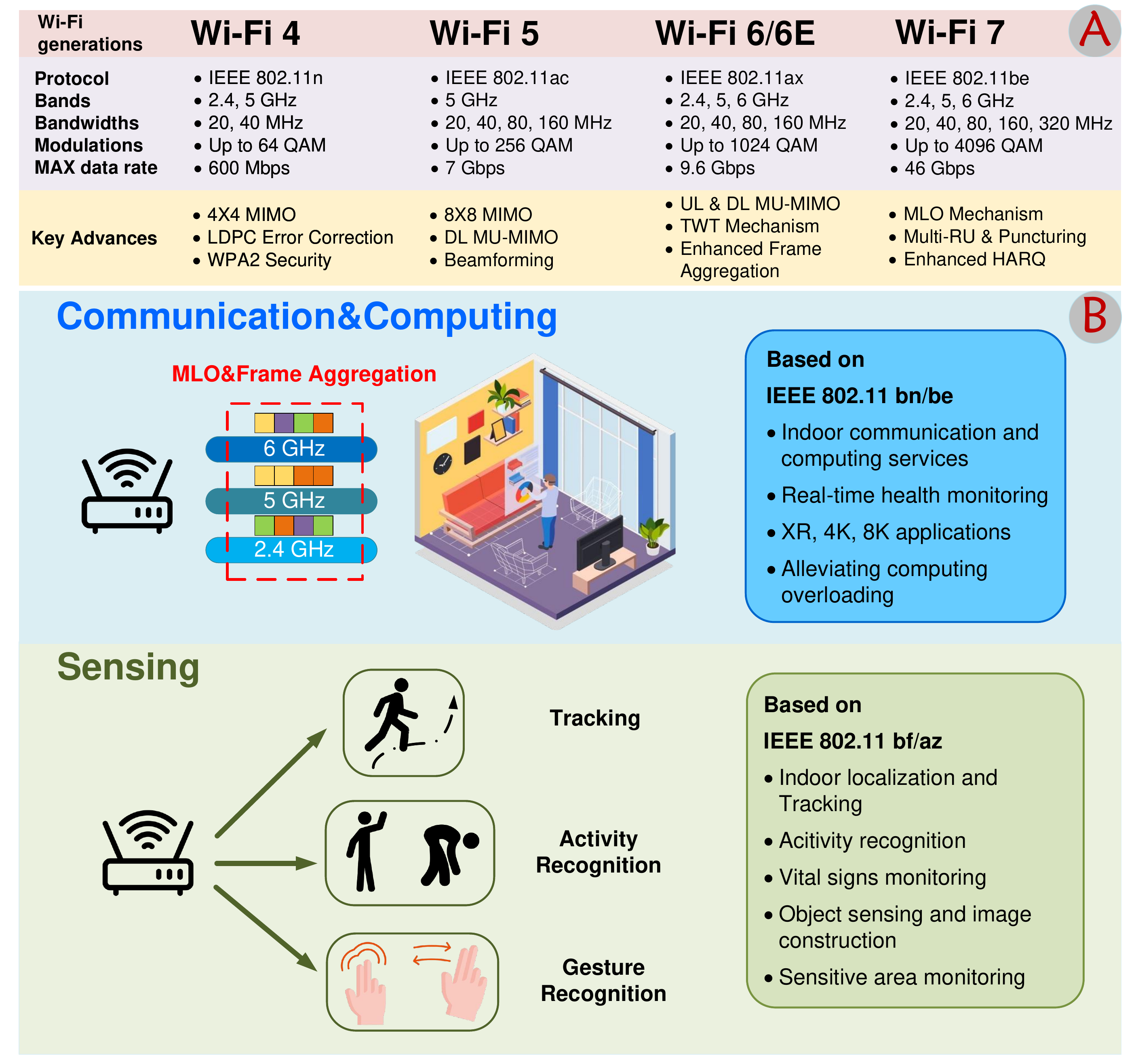}
    \caption{The features of the current Wi-Fi generations and the applications of Wi-Fi networks. Part A compare the main features of Wi-Fi 4 to Wi-Fi 7. Part B summarizes the primary Wi-Fi applications, encompassing communication, sensing, and edge computing. }
    \label{fig1}
\end{figure*}

\section{Overview of Current and Future Wi-Fi Networks}
This section reviews the current Wi-Fi generations, discusses application services like sensing and computing, and illustrates the challenges that future Wi-Fi networks may face.

\subsection{Current Status of Wi-Fi Technology}
Wi-Fi 4\footnote{"IEEE Standard for Information technology-- Local and metropolitan area networks-- Specific requirements-- Part 11: Wireless LAN Medium Access Control (MAC)and Physical Layer (PHY) Specifications Amendment 5: Enhancements for Higher Throughput," in IEEE Std 802.11n-2009 (Amendment to IEEE Std 802.11-2007 as amended by IEEE Std 802.11k-2008, IEEE Std 802.11r-2008, IEEE Std 802.11y-2008, and IEEE Std 802.11w-2009) , vol., no., pp.1-565, 29 Oct. 2009}, the previous and still using Wi-Fi generation, was the first to support dual-band operation in both the 2.4 GHz and 5 GHz bands, achieving a maximum data rate of up to 600 Mbps. Following Wi-Fi 4, Wi-Fi 5\footnote{"IEEE Standard for Information technology-- Telecommunications and information exchange between systemsLocal and metropolitan area networks-- Specific requirements--Part 11: Wireless LAN Medium Access Control (MAC) and Physical Layer (PHY) Specifications--Amendment 4: Enhancements for Very High Throughput for Operation in Bands below 6 GHz.," in IEEE Std 802.11ac-2013 (Amendment to IEEE Std 802.11-2012, as amended by IEEE Std 802.11ae-2012, IEEE Std 802.11aa-2012, and IEEE Std 802.11ad-2012) , vol., no., pp.1-425, 18 Dec. 2013}, based on the IEEE 802.11ac standard, was introduced to the market in 2013 and subsequently enhanced in 2016. Wi-Fi 5 introduced downlink multi-user multiple-input multiple-output (MU-MIMO) and achieve the first gigabit-speed breakthrough. In addition, Wi-Fi 6\footnote{IEEE Standard for Information Technology--Telecommunications and Information Exchange between Systems Local and Metropolitan Area Networks--Specific Requirements Part 11: Wireless LAN Medium Access Control (MAC) and Physical Layer (PHY) Specifications Amendment 1: Enhancements for High-Efficiency WLAN," in IEEE Std 802.11ax-2021 (Amendment to IEEE Std 802.11-2020) , vol., no., pp.1-767, 19 May 2021}, the currently most popular Wi-Fi generation, further advanced wireless networks by operating across the 2.4 GHz and 5 GHz bands, with an extended version, Wi-Fi 6E, adding the 6 GHz band. Wi-Fi 6 introduced several key technologies, including target wake time (TWT), and enhancements to frame aggregation. Frame aggregation technology adopted two types of aggregation mechanisms: aggregated medium access control (MAC) service data unit (A-MSDU) and aggregated MAC protocol data unit (A-MPDU). Notably, Wi-Fi 6 introduced the ability to aggregate frames from multiple traffic identifiers (TIDs) across different quality of service (QoS) traffic classes using A-MPDU, enabling more efficient aggregation for 802.11ax radios.

Following the success of Wi-Fi 6, the IEEE 802.11 organization has turned its attention to the next generation Wi-Fi program, releasing a new amendment standard IEEE 802.11be Extremely High Throughput (EHT), commonly known as Wi-Fi 7\footnote{"IEEE Draft Standard for Information technology--Telecommunications and information exchange between systems Local and metropolitan area networks--Specific requirements - Part 11: Wireless LAN Medium Access Control (MAC) and Physical Layer (PHY) Specifications Amendment: Enhancements for Extremely High Throughput (EHT)," in IEEE P802.11be/D3.0, January 2023 , vol., no., pp.1-999, 1 March 2023.}. Wi-Fi 7 introduces MLO, enhancing the throughput and reducing latency by enabling flexible operation across 2.4 GHz, 5 GHz, and 6 GHz bands. MLO can be categorized into two transmission modes: asynchronous and synchronous \cite{MLO}. The asynchronous mode allows multi-link devices (MLDs) to perform simultaneous transmit and receive (STR) operations, fully capitalizing on the available channels to deliver extremely high throughput. However, it increases power consumption and in-device coexistence (IDC) interference. To mitigate these issues, the synchronous mode has been proposed, where MLDs either transmit or receive at the same time (non-STR), reducing complexity at the cost of some performance.

Although Wi-Fi 7 is still in development, including its features and product advancements, it is essential to prepare for the design of IEEE 802.11bn, known as Wi-Fi 8. Throughout the history of Wi-Fi technologies evolution, it has become noteworthy that simply adopting conventional approaches---such as extending channel bandwidth, increasing modulation coding scheme (MCS) order, and enhancing spatial reuse streams---gradually reaches a performance bottleneck. In consequence, advanced techniques such as machine learning (ML) and GAI are expected to be introduced to overcome these limitations \cite{bn}.

Conventionally, Wi-Fi technologies have primarily been responsible for data transmissions in wireless local area network (WLAN). Recently, the sensing capabilities of Wi-Fi radio signals have gained attention. To achieve this, two relevant protocols have been proposed: Wi-Fi sensing based on IEEE 802.11bf and indoor positioning based on IEEE 802.11az. Leveraging the sensing capabilities, Wi-Fi can improve work efficiency, reduce accidents, and greatly facilitate people's daily lives by enabling activity recognition, object sensing, and localization \cite{bf}. Furthermore, the ubiquitous coverage of Wi-Fi networks has paved the way for mobile edge computing (MEC). Incorporating MEC into Wi-Fi infrastructure can significantly relieve transmission congestion and overloading in cellular network caused by excessive task offloading \cite{EdgeComputing}. Fig. \ref{fig1}(A) presents the features of the current Wi-Fi generations, and Fig. \ref{fig1}(B) presents the main applications.

\subsection{Future Challenges for Wi-Fi Networks}
While Wi-Fi standards have undergone numerous amendments to accommodate emerging applications with rigorous demands, this has led to many new features and a plethora of MAC, and physical (PHY) parameters, resulting in intricate link configuration issues. Accordingly, it is challenging to effectively leverage these features across various scenarios and design high-efficient algorithms to optimize these parameters. We can discuss this from the following two aspects.

\begin{itemize}
    \item{\textbf{Challenges in Wi-Fi Network Analysis and Design:} The complexity of Wi-Fi networks, with numerous parameters across PHY and MAC layers and varying scenario-specific requirements, makes it challenging to analyze the scenarios and choose appropriate performance metrics that accurately reflect their unique characteristics. For example, in a healthcare Internet of Things (H-IoT) system, there are various stringent requirements such as extreme low end-to-end (E2E) delay and bit error rate (BER) for ultra-high reliability, as well as high energy efficiency for wearable devices with short battery life \cite{H-IoT}. To meet these demands, advanced Wi-Fi technologies such as MLO, which enables data transmission across multiple bands, and multi-AP coordination, which allows for coordinated beamforming and joint transmission, are essential for maintaining high-speed, stable connections necessary for consistent vital tracking. Nevertheless, understanding and optimizing these features is complex and time-consuming with traditional manual methods.}
    \item{\textbf{Challenges in Wi-Fi Network Performance Enhancement:} In wireless communication systems, devising strategies to optimize tasks like resource allocation and user scheduling, while accommodating dynamic channels and users' requirements, often incurs high costs with traditional mathematical optimization methods. In Wi-Fi networks, the complexity is further exacerbated by the sheer volume of parameters introduced by constantly evolving protocols (e.g., IEEE 802.11ax and 802.11be). For instance, enhanced frame aggregation requires meticulous design to optimize combinations for different QoS traffic MPDUs, aiming to reduce contention overhead and improve overall network throughput. Additionally, MLO technology significantly increases the complexity of channel access due to the simultaneous operation of multiple links. Currently, some methods, such as deep reinforcement learning (DRL), have been adopted to address some simple Wi-Fi issues. However, they struggle with the vast state and action spaces in future Wi-Fi networks, which are defined by new protocols and numerous access devices, resulting in prohibitively high convergence costs and limited performance. Therefore, resolving these issues effectively remains a significant challenge, and finding optimal or suboptimal solutions is still an area of active research.}
\end{itemize}

To overcome the aforementioned challenges, we resort to LLM and GAI models for analyzing and modeling Wi-Fi networks, as well as enhancing their performance. 

\section{Applications of GAI for Wi-Fi Networks}

In this section, we present various GAI technologies and some related applications. Following that, we elaborate on how LLM assists researchers for Wi-Fi network optimization and how GAI models strength various technologies for Wi-Fi network, as summarized in Table \ref{table1}.

\subsection{GAI Models and Applications}
Compared to traditional AI methods, GAI represents a specific class of AI models with exceptional learning and generalization capabilities. Key technologies involved in GAI include generative adversarial network (GAN), variational autoencoder (VAE), GDM. GAN consists of two neural networks: a generator and a discriminator. The discriminator helps the generator produce data that are indistinguishable from real data. Furthermore, VAE, comprising an encoder and a decoder, compresses input data into a latent space and reconstructs the original data from this latent space, generating new data following the distribution of training data. GAN and VAE models can efficiently extract features from wireless environment due to their ability to learn meaningful latent representations. Hence, these models are often adopted for Wi-Fi localization and predicting potential channel interference \cite{IndoorLocalization,WidebandOperation}. Recently, in wireless communication, GDMs have been increasingly integrated into DRL frameworks to enhance sequential decision-making \cite{GDM_DRL}. GDM starts with a simple noise pattern and gradually refines it through a series of diffusion steps, making small adjustments until the noise resembles the desired output.

As one of the most well-known applications of GAI, LLM has garnered widespread attention. It is an advanced AI system trained on vast amounts of text data, designed to understand and generate human-like text. Nevertheless, LLM often produces misleading or incorrect responses, a phenomenon known as hallucination \cite{AutomatedFeedback}. A promising solution to mitigate hallucination is retrieval-augmented generation (RAG). RAG combines LLM with vast databases from which it can retrieve information, thereby providing more accurate and contextually relevant responses.

The aforementioned models and applications demonstrate robust understanding and modeling capabilities, inspiring us to apply them to Wi-Fi network optimization and performance enhancement.

\subsection{LLM-assisted Wi-Fi Networks Design}
This study focuses on exploiting LLM to help researchers analyze performance requirements, determine metrics, and identify key variables for specific Wi-Fi application scenarios.

When analyzing and modeling various Wi-Fi scenarios, the primary task is to identify the relevant performance metrics that align with the characteristics of the studied scenarios. This task becomes more challenging in complex scenarios that consider multiple metrics simultaneously, making it difficult to derive clear insights. In this context, the LLM can analyze the information provided by users, eliminate trivial factors, and recommend appropriate performance metrics based on their importance. Once performance metrics are set, it is crucial to select appropriate parameters that affect these metrics. Potential parameters span both the PHY and MAC layers and include frame aggregation size, contention window (CW) and operating channels. The LLM with a Wi-Fi knowledge base can help researchers to identify critical variables by considering numerous design requirements. Then based on these chosen variables, the LLM formulates optimization objectives and constraints, thereby improving the design efficiency.

In wireless communication field, there are instances of adopting the paradigm that combine LLM and RAG technologies \cite{LLM_RAG_1,LLM_RAG_satellite}. A pluggable LLM module integrated with a RAG module is proposed to promote efficient network management in  \cite{LLM_RAG_1}. A similar framework is also utilized to achieve reasonable problem formulations for satellite communication networks  \cite{LLM_RAG_satellite}. Simulation results demonstrate that the LLM and RAG framework can provide more accurate and efficient modeling compared to traditional manual methods. Accordingly, we aim to utilize this framework to efficiently model future Wi-Fi networks, which are becoming increasingly complex.

\subsection{GAI-assisted Wi-Fi Networks Performance Enhancement}
\begin{table*}[htp]
	\centering  % 显示位置为中间
	\caption{Applications of GAI for Wi-Fi Networks.}  % 表格标题
	\label{table1}  % 用于索引表格的标签
	%字母的个数对应列数，|代表分割线
	% l代表左对齐，c代表居中，r代表右对齐
    % p代表上对齐，m代表居中，b代表底对齐
    % l和p无法连用
	\begin{tabular}{|m{2.5cm}<{\centering}|c|c|c|c|}  
		\hline  % 表格的横线
		% &  &  & \\[-6pt]   %可以避免文字偏上来调整文字与上边界的距离
		\textbf{Research Directions} & \rule{0pt}{3ex} \textbf{Challenges}&\textbf{Application Examples} \rule[-1.5ex]{0pt}{0pt}&\textbf{Advantages}&\textbf{References} \\  % 表格中的内容，用&分开，\\表示下一行
		\hline
        Wi-Fi Network Parameter Design
        &\multicolumn{1}{m{3cm}|}{Accurate model design including proper performance metrics and parameters selection for various complex scenarios.} 
        &\multicolumn{1}{m{5cm}|}{
            \begin{itemize}[leftmargin=5pt, after=\vspace{-\baselineskip}]
                \item{Assistance in selecting suitable performance metrics that reflect the critical characteristics of studied scenario.} 
                \item{Analyzing design requirements to select key parameters that are closed related to research objective.}
            \end{itemize}} 
        &\multicolumn{1}{m{4cm}|}{
            \begin{itemize}[leftmargin=5pt, after=\vspace{-\baselineskip}]
                \item{Remarkable interaction and understanding abilities for performance analysis.} 
                \item{Strong analytical ability to derive clear insights from complex design requirements.}
            \end{itemize}}
        &\multicolumn{1}{m{1.5cm}|}{\makecell{\cite{LLM_RAG_1, LLM_RAG_satellite}}}  \\% \makecell 居中
		\hline
        Wi-Fi Network Performance Enhancement
        &\multicolumn{1}{m{3cm}|}{Solving intractable network optimization which incur high costs, especially when using traditional rule-based methods.}
        &\multicolumn{1}{m{5cm}|}{
            \begin{itemize}[leftmargin=5pt, after=\vspace{-\baselineskip}]
                \item{Adopt the GAN to predict interference for optimal channel bonding configuration.} 
                \item{Leverage the VAE to predict optimal beam sector pairs for adaptive beam-tracking.}
                \item{GAI models can optimize frame size based on specific channel conditions.}
                \item{GAI models can optimize traffic allocation based on link occupancy.}
            \end{itemize}}
        &\multicolumn{1}{m{4cm}|}{
            \begin{itemize}[leftmargin=5pt, after=\vspace{-\baselineskip}]
                \item{A more efficient solution to the intractable problem of improving network performance.} 
                \item{Intelligently optimizing parameters to adaptively align with wireless link.}
            \end{itemize}}
        &\multicolumn{1}{m{1.5cm}|}{\makecell{\cite{WidebandOperation, Beamforming}}}  \\ % \makecell 居中
        \hline
        Other Wi-Fi-based Applications
        &\multicolumn{1}{m{3cm}|}{Emergence of other Wi-Fi applications with distinct performance metrics and requirements that entails more effective and customized approaches.}
        &\multicolumn{1}{m{5cm}|}{
            \begin{itemize}[leftmargin=5pt, after=\vspace{-\baselineskip}]
                \item{Adopt the GDM to design a resource allocation strategy that balances task offloading and communication performance.}
                \item{Leverage the VAE to capture data features and enhance indoor localization accuracy.} 
                \item{The GAN can generate high-quality wireless data, addressing data shortages in network security.}
            \end{itemize}}
        &\multicolumn{1}{m{4cm}|}{
            \begin{itemize}[leftmargin=5pt, after=\vspace{-\baselineskip}]
                \item{More powerful generalization ability to make contributions to various Wi-Fi applications.} 
                \item{Better performing in tradition AI-based applications areas due to its better ability to capture the distribution of the complex and high-dimensional data.}
            \end{itemize}}  
        &\multicolumn{1}{m{1.5cm}|}{\makecell{\cite{IndoorLocalization, EdgeComputing, NetworkSecurity}}}  \\ % \makecell 居中
        \hline
	\end{tabular}
\end{table*}
To date, the development of Wi-Fi protocols has spanned multiple generations, introducing features such as enhanced frame aggregation in IEEE 802.11ax and MLO in IEEE 802.11be. Achieving optimal performance in Wi-Fi systems typically necessitates the utilization of these new features and the fine-tuning of numerous parameters. Traditional rule-based methods are restricted by human effort and expertise, making them time-consuming and often inadequate for finding the best solutions. In this case, GAI models can aid researchers in seeking optimal solutions and enhancing network performance according to various requirements.

\begin{enumerate}
    \item{\textbf{Key Technologies in PHY Layer:} The PHY layer is responsible for the actual transmission of data over wireless medium. Key PHY layer technologies include:}
    \begin{enumerate}
        \item{\textit{Wideband Operation:} Channel bonding, through the aggregation of multiple channels, increases available bandwidth for data transmission, thus augmenting data throughput and overall network performance. However, it is challenging to optimize channel allocation and bonding configurations to mitigate potential contention, especially regarding hidden channel interference issues. Since GAI models have powerful learning and generalization capabilities, it can fully capture the scenario characteristics and predict potential channel interference, thereby ensuring optimal channel configurations. A novel generative network framework known as the Metropolis-Hastings generative adversarial network (MH-GAN) was proposed in \cite{WidebandOperation}. This framework effectively forecasts hidden channel interference from neighboring basic service sets (BSSs). Performance analysis conducted on a testbed deployment demonstrates that the proposed model significantly improves throughput compared to the baseline approaches and the standard IEEE 802.11 operations.}
        \item{\textit{Beamforming:} While beamforming allows Wi-Fi devices concentrating the signal towards the intended receiver, greatly augmenting the transmission rate, determining optimal beam sector pairs entails considerable complexity and overhead. To cope with this challenge, GAI models can learn the environmental dynamics and replace the time-consuming beam sweeping procedure with the prediction of optimal beam sector pairs. For example, a VAE-based method was investigated for designing adaptive beam-tracking with low overhead \cite{Beamforming}. A two-timescale framework was introduced: over a long timescale, a deep recurrent variational autoencoder (DR-VAE) learns beam dynamics and enable predictive beam-tracking; over a short timescale, based on the strongest beam pair predicted by DR-VAE, an adaptive beam-training procedure is optimized through point-based value iteration (PBVI) method. Numerical results show that this approach improves the spectral efficiency and outperforms other baseline algorithms.}
    \end{enumerate}
    \item{{\textbf{Key Technologies in MAC Layer:}} he MAC layer handles access control, data framing, and error handling, ensuring efficient and reliable data transmission. Key MAC layer technologies include:}
    \begin{enumerate}
        \item{\textit{Frame Aggregation:} It is vital to select the frame length that matches with dynamic channel. While larger frames can reduce the influence of the overhead, they are more prone to transmission errors under poor channel conditions, hence demanding deliberating on the ratio of useful transmitted data to overhead. Considering the powerful capability of capturing peculiarities of the wireless channel, the GAI models can be employed to address this trade-off by deriving the optimum frame length tailored to specific channel conditions. For example, in a Wi-Fi network, the access point (AP) can calculate the packet loss rate by monitoring the ACK feedback from stations and measure the channel idle time. The GAI models analyze this collected information and instruct stations to adjust their frame length settings, thereby maximizing the system efficiency.}
        \item{\textit{Multi-link Operation:} In Wi-Fi 7, while MLO introduces an additional degree of freedom by using multiple bands, it also entails complexity in channel selection and traffic allocation across different links. An improper configuration can cause severe congestion instead of improved throughput. To address this, GAI models can analyze the occupancy of 2.4 GHz, 5 GHz and 6 GHz bands and intelligently allocate traffic to the most suitable channels in each band, thereby relieving transmission congestion and increasing the chances of successful transmission. Moreover, in the non-STR mode, inefficient resource scheduling can lead a marked delay cost due to the necessary frame synchronization process. To mitigate this, GAI models can be employed to optimize frame transmission, reducing synchronization overhead and thus enhancing communication quality.}
    \end{enumerate}
\end{enumerate}

\subsection{Roles of GAI in Other Wi-Fi-based Applications}
Besides Wi-Fi networks design and performance optimization, GAI can greatly contribute to areas like computing, sensing, and security. The following examples highlight GAI's impact in these domains.

\begin{figure*}[htp]
    \centering
    \includegraphics[width=1\textwidth]{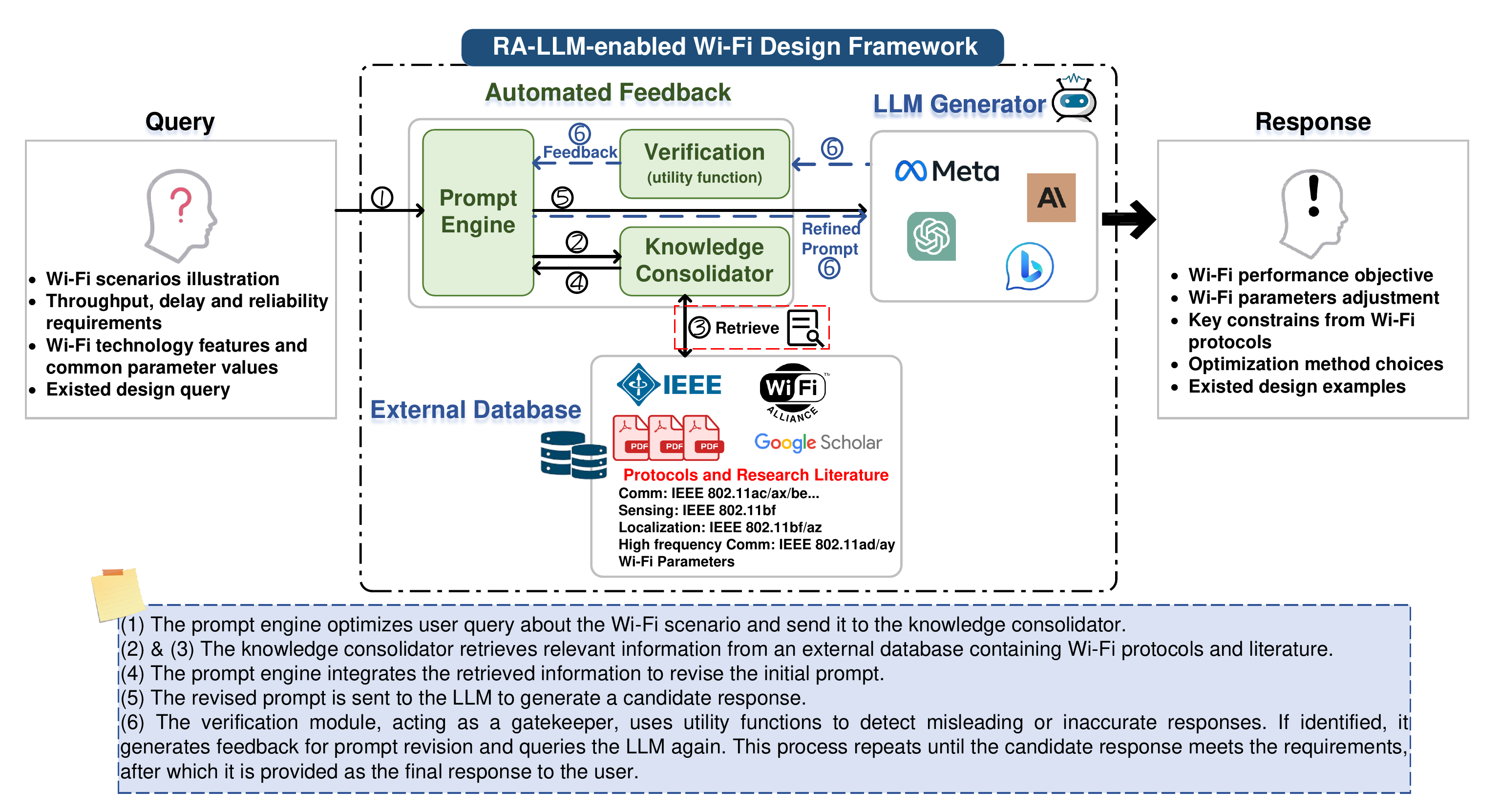}
    \caption{The proposed framework for Wi-Fi network optimization.}
    \label{fig2}
\end{figure*}

\begin{enumerate}
    \item{\textbf{Edge Computing:} Edge computing applications heavily depend on high throughput and reliability provided by IEEE 802.11 communication protocols, such as IEEE 802.11be and 802.11bn. Therefore, it is crucial to balance communication and computation offloading to meet task requirements. One effective approach to deal with these issues is to utilize the analytical and modeling capabilities of GAI models. A hybrid algorithm framework was proposed to efficiently resolve communication and offloading coexistence issues \cite{EdgeComputing}. This framework combines a GDM-based RL algorithm, called diffusion twin delayed DDPG (DTD3), with a resource allocation scheme. It first makes offloading decisions on the basis of observed task properties and MEC server resources, then allocates communication and computational resources to fulfill task requirements. Simulation results demonstrate that the proposed algorithm framework significantly reduces task processing latency and total energy consumption.}
    \item{\textbf{Indoor Localization Service:} Wi-Fi protocols like IEEE 802.11bf and 802.11az provide a solid foundation for indoor localization. Moreover, it is important to highlight that deep learning frameworks have greatly advanced Wi-Fi-based localization accuracy but necessitate substantial labeled data, which is costly to obtain. However, GAI models, such as VAE, offer a solution by mapping input data into a latent space that captures essential features, allowing for effective generalization even with limited annotated Wi-Fi observations. For example, a VAE-based semi-supervised method proposed in \cite{IndoorLocalization}  utilizes a hybrid dataset of a small amount of labeled data and a large unlabeled dataset to develop an accurate predictor for localization. The VAE treats the unseen label as the latent variable and updates itself according to the marginal likelihood from both labeled and unlabeled data. The proposed solution is validated on public dataset, demonstrating the model's data efficiency and its superior performance compared to state-of-the-art approaches.}
    \item{\textbf{Data Augmentation for Network Security:} As a type of edge network, Wi-Fi networks handle large volumes of data transmission, often involving sensitive user information, which makes them vulnerable to security breaches. Unfortunately, Wi-Fi network is less likely to implement a complex intrusion detection mechanism enabled by massive available data. To address this, GAI for data augmentation can be applied  in scenarios where acquiring labeled data is time-consuming and costly. A GAN architecture was introduced to generate high-quality, labeled wireless data using a limited amount of real data \cite{NetworkSecurity}. The authors then trained a convolutional neural network (CNN) on a hybrid dataset containing both real and synthetic data to classify wireless nodes for a wireless physical layer authentication model. Simulation results indicate that the inclusion of the synthetic dataset improves classification accuracy compared to using only the original dataset.}
\end{enumerate}

In summary, GAI is poised to bring a paradigm shift to future Wi-Fi networks by simplifying design processes and enhancing performance. Specifically, the LLM, with rich knowledge and powerful reasoning abilities, can assist researchers in mathematical modeling for complex Wi-Fi scenarios. Meanwhile, GAI models serve as invaluable tools for optimizing intricate Wi-Fi link configurations, which are complicated by numerous novel features and parameters, and for expanding the technical application of Wi-Fi in diverse scenarios.

\section{Case Study}
This section presents a use case in which we adopt an LLM framework to assist in mathematical modeling of Wi-Fi scenarios and adopt a GAI model to offer solutions for Wi-Fi network optimization. Finally, we validate the solution using the NS-3 platform.

\subsection{LLM-enabled Wi-Fi Optimization Framework}
\begin{figure*}[htp]
    \centering
    \includegraphics[width=1\textwidth]{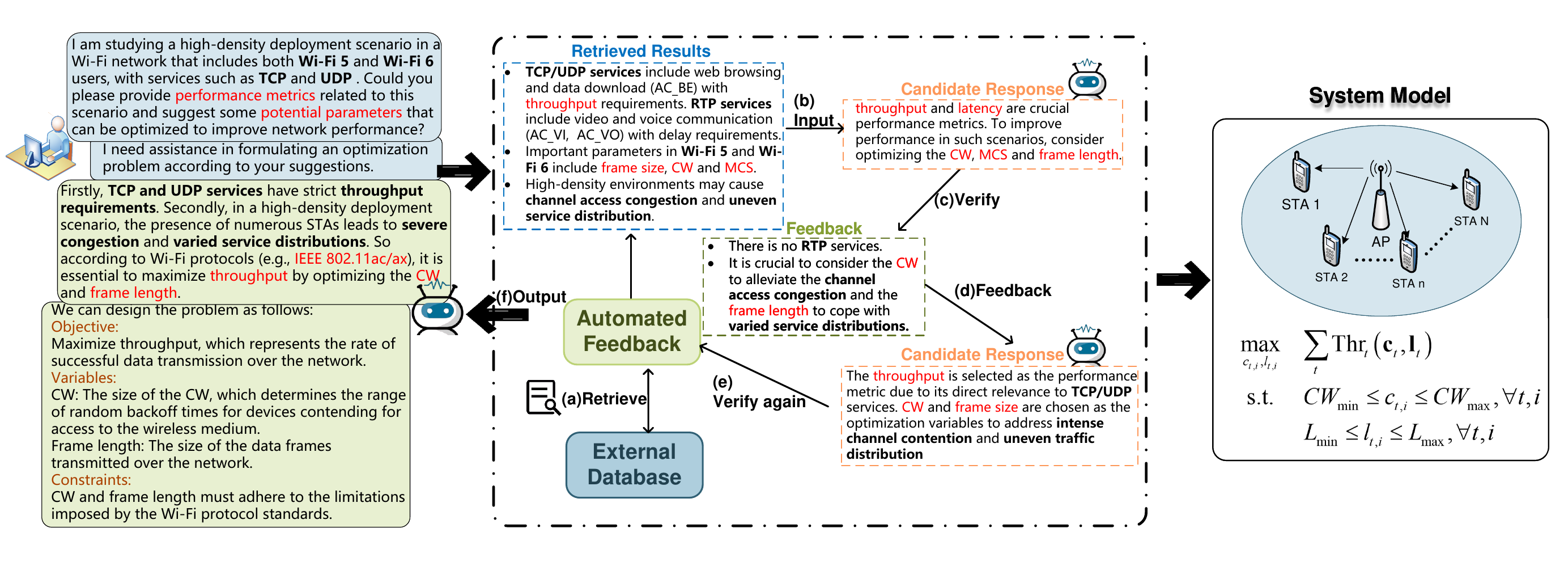}
    \caption{The case to use the proposed framework to assist in formulating problem.}
    \label{fig3}
\end{figure*}
As shown in Fig. \ref{fig2}, we propose a retrieval-augmented LLM (RA-LLM)-enabled Wi-Fi optimization framework. the framework is composed of three parts: an LLM generator, an external database module, and an automated feedback module. Although many studies have demonstrated that the paradigm combining LLM and RAG technologies is highly beneficial for researchers analyzing wireless communication systems in various scenarios, the automated feedback mechanism further reduces LLMs' hallucinations without sacrificing the fluency and informativeness of its responses \cite{AutomatedFeedback}. Therefore, we integrate this module into our framework which is specialized in Wi-Fi networks optimization. 

Specifically, the automated feedback module consists of three components, a prompt engine, a knowledge consolidator, and a verification module. The prompt engine optimizes the received information (e.g., user queries, verification feedback, retrieved data) and sends it to the corresponding module, such as the consolidator and generator. The knowledge consolidator module receives the initial prompt and retrieves relevant information from the external knowledge base. It connects the retrieved information to its corresponding descriptions in IEEE 802.11 protocols or related literature, creating a shortlist of the most pertinent evidence chains for the queries. After integrating the retrieved information with the initial prompt, the consolidator sends it back to the prompt engine for further revision. The verification module then acts as a gatekeeper, leveraging utility functions to identify if the LLM generates misleading or inaccurate responses. If so, this module will generate the informative feedback messages for prompt revision and queries LLM again. This process repeats until the candidate response meets the required criteria, at which point it is provided as the final response to the user. The verification module can employ two types of utility function, model-based utility functions and rule-based utility functions. The former needs to be trained on label datasets, while the latter is usually based on heuristics or programmed functions \cite{AutomatedFeedback}.

\subsection{Wi-Fi Network Performance Enhancement in High-density Deployment Scenarios}

\textbf{Problem Formulation:}  With the surge in Wi-Fi devices worldwide, optimizing network performance in high-density deployment scenarios has become a critical research topic \cite{dengcailian}. For this reason, we design an optimization model for a Wi-Fi network that includes one AP and numerous stations, comprising both Wi-Fi 5 and Wi-Fi 6 users, randomly distributed around the AP, and running TCP and UDP services. The LLM, by retrieving relevant Wi-Fi knowledge from external databases, identifies throughput as a key performance metric for these services. In addition, high-density environments lead to channel congestion and uneven service distribution, with the Binary Exponential Backoff (BEB) mechanism limiting performance due to its inflexible CW adjustment. Besides, intelligent configuration of frame size can significantly alleviate the traffic stress caused by multiple stations. Therefore, through iterative refinement using an automated feedback module, the LLM formulates an optimization model that maximizes the total throughput by jointly optimizing CW and frame length, as illustrated in Fig. \ref{fig3}. 

\begin{figure*}[htp]
    \centering
    \includegraphics[width=1\textwidth]{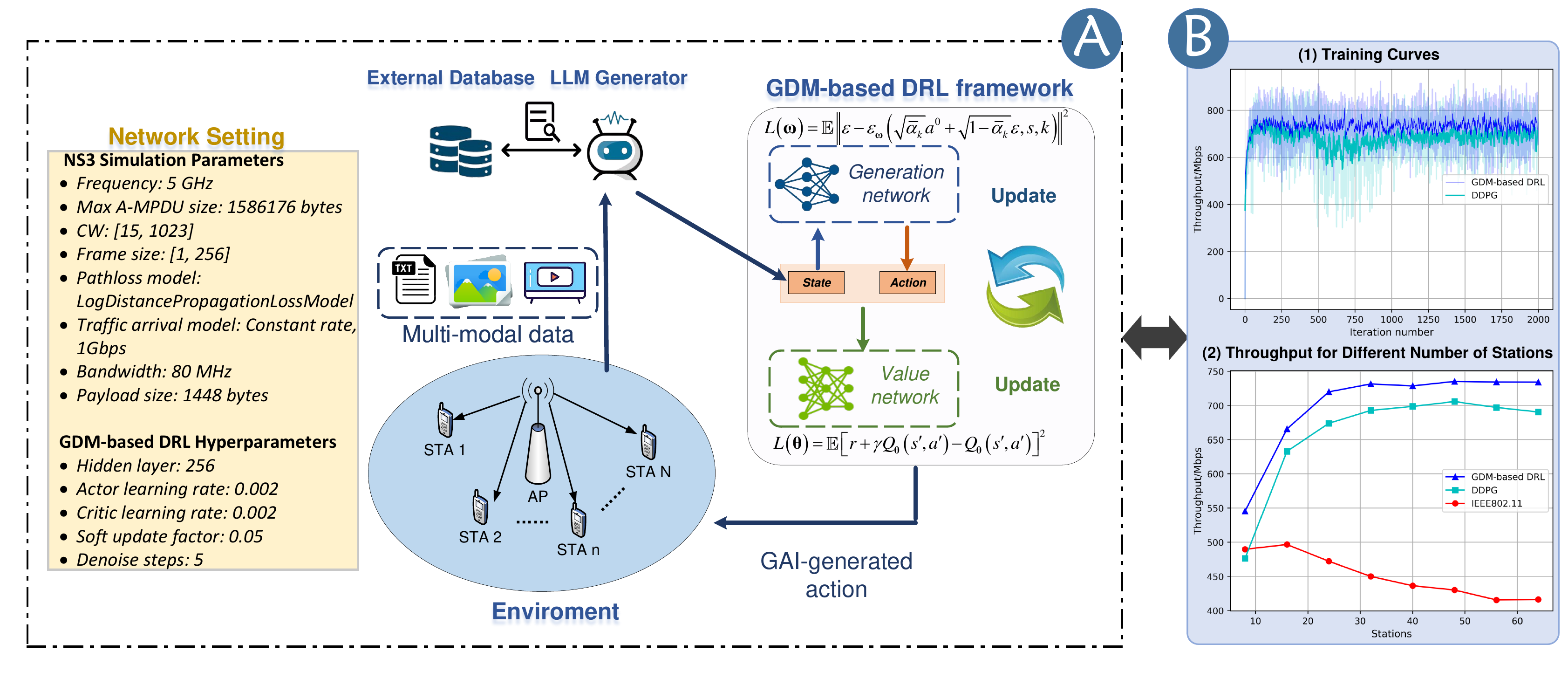}
    \caption{GDM assisted throughput maximization for high-density deployment scenarios. We compare the convergence speed of GDM and DDPG, followed by throughput comparisons across different baselines and numbers of stations.}
    \label{fig4}
\end{figure*}
\textbf{Performance Evaluation:} As illustrated in Fig. \ref{fig4}(A), we adopt a GDM-based DRL framework that optimizes variables through continuous interaction with the environment. Compared with other GAI models, GDM excels at modeling complex data distribution and its inherent denoising process enhances exploration, avoiding local optima and outperforming traditional DRL methods \cite{GDM_DRL}. Additionally, information in a wireless environment comes not only from RF signals but also from multi-modal sources such as LiDAR, camera images and GPS. By employing the LLM to handle these multi-modal observations, non-RF modalities can effectively complement the state features, enabling the DRL framework to fully capture the network conditions and make informed decisions.

The proposed solution is validated on the NS-3 platform, which provides a rich set of models for simulating and analyzing the performance of various network protocols. As shown in Fig. \ref{fig4}(B.1), the proposed algorithm achieves higher throughput and exhibits a more stable training process compared to the DDPG algorithm. This is because GDM has a superior capability for exploring the different state-action pairs and capturing complex patterns in the dynamic wireless environment, thus avoiding local optima. Fig. \ref{fig4}(B.2) shows that as the number of stations increases, the proposed algorithm ensures monotonic growth in throughput, outperforming the DDPG algorithm. For the baseline leveraging 802.11 Wi-Fi standard, the throughput initially increases but degrades as the number of stations continues to grow. This is due to that a large number of stations exceed the capacity of the Wi-Fi network, and the MAC layer adjusts the CW to a small value to avoid fierce competition, resulting in a decline in total throughput. It is important to highlighted that, even under network saturation, our algorithm maintains a stable throughput without drastic decline by intelligently setting CW and frame length. Therefore, the simulation results demonstrate the superiority of our proposed algorithm for high-density deployment scenarios.

\section{Future Directions}
\subsection{Interference Mitigation}
Interference poses a fundamental challenge in Wi-Fi networks using unlicensed spectrum. GAI offers a viable solution by capturing spatiotemporal properties of interference and predicting their occurrences. By harnessing GAI's data analysis, prediction, and optimization capabilities, Wi-Fi networks can achieve more efficient and reliable performance in these environments.

\subsection{Privacy Protection}
Deploying GAI models in Wi-Fi networks necessitate training on the extensive traffic datasets, raising serious privacy concerns. For this reason, future research should prioritize privacy protection. Federated learning offers a privacy-preserving alternative by leveraging local computation and secure centralized model aggregation, thereby serving as a promising solution to these issues. 

\subsection{Multi-modal Data Processing and Feature Analysis}
Although multi-modal data can be utilized to enhance communication system performance, simply combining these data sources is insufficient for real-world applications. Leveraging GAI models and semantic methods can effectively extract features aligned with specific objectives and eliminate redundancy. While LLMs have shown potential in handling multi-modal data, further research is needed to refine these models (e.g., structure, fine-tuning methods) to better accommodate the complexities of wireless environments. 

\section{Conclusion}
In this article, we have investigated the GAI's applications in Wi-Fi networks. We began by introducing the current using Wi-Fi generations and highlighting the potential challenges in future Wi-Fi systems. Subsequently, we detailed the typical GAI models and their support in Wi-Fi network optimization, performance enhancement, and other Wi-Fi-based applications. Moreover, we presented a case study proposing a novel LLM-enabled Wi-Fi design framework for reasonable problem formulation, followed by the use of the GDM to maximize network throughput. Numerical results demonstrate the superiority of our proposed algorithm in high-density deployment scenarios. Finally, some future directions are discussed.

% \section*{Acknowledgments}
% This should be a simple paragraph before the References to thank those individuals and institutions who have supported your work on this article.

\bibliographystyle{IEEEtran}
\bibliography{reference}

% \vspace{-33pt}
% Use $\backslash${\tt{begin\{IEEEbiographynophoto\}}} and author name as the argument followed by the biography text.

\vfill

\end{document}